\documentstyle[aaspp4,tighten,flushrt]{article}

\newcommand{\beq}{\begin{equation}}
\newcommand{\eeq}{\end{equation}}
\newcommand{\beqa}{\begin{eqnarray}}
\newcommand{\eeqa}{\end{eqnarray}}

\newcommand{\lexp}{\mathop{\langle}}
\newcommand{\cbar}{\mathop{|}}
\newcommand{\rexp}{\mathop{\rangle}}
\newcommand{\rexpc}{\mathop{\rangle_c}}

\def\eq#1{equation~(\ref{#1})}

\def\d{\delta}
\def\D{\Delta}
\def\Dk{\Delta \kappa}
\def\dD{\delta_{\rm D}}

\def\nt{{\tilde n}}
\def\Ft{{\tilde F}}
\def\Dstar{{D^\ast}}
\def\eff{{\rm eff}}
\def\N{{\cal N}}
\def\Nt{\tilde{\cal N}}
\def\P{{\cal P}}
\def\B{{\cal B}}
\def\DD{{\cal D}}

\font\BF=cmmib10 
\font\BFs=cmmib10 scaled 833
\def\k{\hbox{\BF k}}
\def\q{\hbox{\BF q}}
\def\r{\hbox{\BF r}}
\def\x{\hbox{\BF x}}

\def\ks{\hbox{\BFs k}}
\def\qs{\hbox{\BFs q}}
\def\rs{\hbox{\BFs r}}
\def\xs{\hbox{\BFs x}}
\def\nhat{\hat{\hbox{\BF n}}}

\def\bth {\hbox{\BF  \char'022}} 
\def\bkp {\hbox{\BF  \char'024}} 
\def\bths{\hbox{\BFs \char'022}}
\def\bkps{\hbox{\BFs \char'024}}

\begin{document}

\title{Projection and Galaxy Clustering Fourier Spectra}

\author{J. N. Fry \& David Thomas}
\affil{Department of Physics, University of Florida, 
Gainesville, FL  32611-8440}
\authoremail{fry@phys.ufl.edu, davet@oddjob.uchicago.edu}

\begin{abstract}

Second order perturbation theory predicts a specific dependence 
of the bispectrum, or three-point correlation function in the 
Fourier transform domain, on the shape of the configuration of 
its three wave vector arguments, which can be taken as a signature 
of structure formed by gravitational instability.
Comparing this known dependence on configuration shape with the 
weak shape dependence of the galaxy bispectrum has been suggested 
as an indication of bias in the galaxy distribution.
However, to interpret results obtained from projected catalogs, 
we must first understand the effects of projection on this 
shape dependence.

We present expressions for the projected power spectrum and 
bispectrum in both Cartesian and spherical geometries, 
and we examine the effects of projection on the predicted  
bispectrum with particular attention to the dependence on 
configuration shape.
Except for an overall numerical factor, for Cartesian projection 
with characteristic depth $ \Dstar $ there is little effect 
on the shape dependence of the bispectrum for wavelengths small 
compared to $ \Dstar $ or projected wavenumbers $ q \Dstar \gg 1 $.
For angular projection, a scaling law is found for spherical harmonic 
index $ \ell \gg 1 $, but there is always a mixing of scales over 
the range of the selection function.
For large $ \ell $ it is sufficient to examine a small portion 
of the sky.

\end{abstract}

\keywords{large-scale structure of universe}

\section{Introduction}

The behavior of cosmological density fluctuations in perturbation 
theory is becoming a mature, well-understood subject, 
allowing analytic calculations of the expected behavior of 
correlation functions to high order (Fry 1984; Goroff et al.~1986; 
Bernardeau 1992; Jain \& Bertschinger 1994; 
Scoccimarro \& Frieman 1996a,b; Scoccimarro et al.~1998).
Starting with Gaussian initial conditions at early times, the first 
nonvanishing contribution to the $n$-point correlation function 
$ \xi_n $ requires knowledge of perturbations to order $ n-1 $.
To leading order irreducible point or windowed moments of 
density follow the hierarchical pattern, 
\beq
\xi_n = \lexp \d^n \rexpc = S_n \, \xi_2^{n-1} , 
\eeq
first found in observations, where the full galaxy three-point 
correlation function $ \zeta $ can be written as a sum of products 
of two-point functions $ \xi $ as (Groth \& Peebles 1977) 
\beq
\zeta_{123} = Q_{123} \, (\xi_{12} \xi_{13} + \xi_{12} \xi_{23} 
+ \xi_{13} \xi_{23}) . \label{zeta123} 
\eeq
In observations, of both the three-point function in space and 
the bispectrum in the Fourier transform domain (Fry \& Seldner 1982), 
the dimensionless amplitude $Q$ depends only weakly 
on overall scale or on configuration shape.
In perturbation theory, the reduced three-point function amplitude, 
or the corresponding bispectrum amplitude in Fourier space
(see eq.~[\ref{Q123}] below), is not a constant but a function of 
the lengths of the triangle or of the three wave vectors that make 
up the configuration (Fry 1984; Bouchet et al. 1995), 
a function that varies with the power spectrum index $n$ but depends 
only weakly on cosmological parameters such as the fraction of 
critical density $ \Omega_0 $ and the cosmological constant $ \Lambda $ 
(Bouchet et al.~1992; Bouchet et al.~1995; Scoccimarro et al.~1998).
The apparent disagreement between theory and observation can be 
understood as an effect of bias.
A nonlinear but local bias in the galaxy distribution, 
with fractional contrast in the galaxy number density determined 
as a function of the local mass density contrast, 
$ \d_g = \sum b_k \d^k /k! \approx b \d + \frac12 b_2 \d^2 $, 
gives a galaxy three-point amplitude that depends on the first 
two coefficients in the expansion (Fry \& Gazta\~{n}aga 1993), 
\beq
Q_g = {1 \over b} Q_{123} + {b_2 \over b^2} , \label{Q_g}
\eeq
where $ b = b_1 $ is the usual linear bias factor.
Matarrese, Verde, \& Heavens (1997) have estimated the expected 
uncertainties and analyzed numerical simulations to investigate 
the practicality of applying the method to data, and 
have concluded that the method works well in the unbiased case, 
with likelihood centered around $ b=1 $ and $ b_2 = 0 $.
The local bias model may be an oversimplification, but represents 
the minimal nonlinear degree of freedom in any nonlinear 
galaxy formation model.
Comparison of the observed galaxy three-point amplitude in the Lick 
catalog with that expected in perturbation theory using \eq{Q_g} 
provides an estimate for $b$ that is larger than expected from 
other methods (Fry 1994).

Any measurements that provide information about bias are important 
for our ability to infer properties of the cosmological mass 
distribution from galaxy observations.
However, many of the largest galaxy compilations, including the Lick 
catalog, are angular positions only, and an important question is 
whether angular projection has a significant effect on the results.
For angular correlation functions and moments this has been studied 
for some time 
(Limber 1953; Rubin 1954; Groth \& Peebles 1977; Kaiser 1992; 
Pollo 1997; Gazta\~{n}aga \& Bernardeau 1998).
In this paper we compute the effects of projection in the 
transform domain, on the power spectrum and bispectrum.
In the following, in Section 2 we first compute a simple 
Cartesian projection, where we can easily observe the approach 
to the asymptotic small and large scale results, 
and present numerical results in Section 3.
Following this, we examine angular projection in Section 4, 
over the full sky and also over a small patch.
Section 5 contains a final discussion.

\section{Cartesian Projection}

For simplicity, we first present results for projection along 
a Cartesian direction, denoted $z$.
Cartesian projection misses an important scale mixing found below 
in projection at fixed angular positions, but it is simpler both 
to compute and to understand, and thus provides a useful guide 
in understanding later results.
The projected number density of galaxies $n_p$ is the space density 
$ n(x,y,z) $, summed over $z$, weighted by a selection function $ F(z) $ 
that gives the probability that an object at $z$ is included, 
\beq
n_p (x,y) = \int dz \, F(z) \, n(x,y,z) .
\label{n_p}
\eeq
For convenience, although this is not necessary we take 
$F$ to scale with a typical depth $ \Dstar $, 
$ F = F(z/\Dstar) $ (Peebles 1980).
In an attempt to separate projected and unprojected quantities, 
in what follows we will write projected functions with 
a subscript $p$: $ n_p $, $ \xi_p $, etc.
To distinguish positions in the two-dimensional projected space from 
those in the full three dimensions, we will denote the former 
as $\x$ and the latter as $\r$; similarly, wave vectors in the 
two-dimensional space we will denote $\q$ and in three dimensions 
as $\k$.
It will occasionally be necessary to refer to a three-vector 
made from a two-vector and a third component; 
$ \r = (\x,z) $, $ \k = (\q,k) $, etc.; 
$ \k = \q $ means $ \k = (\q,0) $.
Thus, equation (\ref{n_p}) can equivalently be written
$ n_p (\x) = \int dz \, F(z) \, n(\r) $.

We wish to study statistics of the projected density.
In space, we write $ \lexp n(\r) \rexp = \bar n $, 
and the two-point correlation function $ \xi $ and 
the three-point function $ \zeta $ are defined by
\beq
\lexp n(\r_1) n(\r_2) \rexp = \bar n^2 \left[1+\xi(\r_{12})\right] 
\eeq
and
\beq
\lexp n(\r_1) n(\r_2) n(\r_3) \rexp = \bar n^3 
\left[ 1 + \xi(\r_{12}) + \xi(\r_{23}) + \xi(\r_{31}) 
+ \zeta(\r_1,\r_2,\r_3) \right] , 
\eeq
where angle brackets $ \lexp\cdots\rexp $ indicates an ensemble 
average, equivalent by homogeneity to an average over space, 
and $\r_{12}$ is shorthand for $\r_1-\r_2$.
Cosmological symmetries imply that $\xi$ and $\zeta$ depend 
only on distances between points.

Similar expressions hold for projected functions.
The average projected density is 
$ \bar n_p = \int dz \, F(z) \, \bar n $, 
while from equation (\ref{n_p}) we can express moments of $ n_p $ 
as integrals over the spatial correlation functions, 
\beq
\bar n_p^2 \, \xi_p(\x_{12}) = \bar n^2 \int dz_1 dz_2 
\, F(z_1 )F(z_2) \, \xi(|\r_{12}| ) , 
\eeq
\beq
\bar n_p^3 \, \zeta_p (\x_1, \x_2, \x_3) = 
\bar n^3 \int dz_1  dz_2 dz_3 \, F(z_1) F(z_2) F(z_3) \,
\zeta (\r_1, \r_2, \r_3 ) .
\eeq
These are equivalent to relations for projected angular functions 
(Limber 1953; Peebles 1980).

Because correlations depend only on separations, it is natural to 
change variables from $ z_1 $, $ z_2 $ to the appropriate mean 
$ z = \frac12 (z_1 + z_2) $ or $ z = \frac13 (z_1 + z_2 + z_3) $
and differences $ z_{12} = z_1 - z_2 $.
In the approximation that correlations are small except for small 
separations, we can write 
\beq
\xi_p(\x_{12}) = {\int dz \, F^2(z) \, \over [\int dz \, F(z)]^2} 
\, \int dz_{12} \, \xi( |\r_{12}| )  \label{xi_p} 
\eeq
\beq
\zeta_p( \x_{12}, \x_{13}) 
= {\int dz \, F^3(z) \, \over [\int dz \, F(z)]^3} 
\, \int dz_{12} dz_{13} \, \zeta (\r_{12} , \r_{13} ) \label{zeta_p}
\eeq
From these, it is apparent that if the spatial three-point function is 
hierarchical with constant $Q$ (eq.~[\ref{zeta123}]), 
then so is the projected three-point function, with 
\beq
{Q_p \over Q} = {[\int dz \, F^3(z) ][\int dz \, F(z)] 
\over [\int dz \, F^2(z)]^2} . 
\eeq
This result is independent of the normalization of $F$.
By the Schwarz inequality, $ Q_p/Q \geq 1 $.
The equivalent result was obtained in the angular case by 
Fry \& Seldner (1982).

Our goal now is to compute projected moments in the Fourier domain.
In Fourier transform space we have the amplitude 
\begin{eqnarray}
\nt(\k) &=& \int d^3 r \, n(\r) e^{-i\ks\cdot\rs} , \label{FT} \\
n(\r) &=& \int {d^3 k \over (2\pi)^3} \, \nt (\k) e^{i\ks\cdot\rs} .
\end{eqnarray}
The mean density contributes $ (2\pi)^3 \dD(\k) \bar n $ 
to $ \nt(\k) $.
Moments of the Fourier amplitude for $ \k \neq 0 $ give the power 
spectrum  $ P(k) $ and the bispectrum $ B(\k_1, \k_2, \k_3) $ 
(the Fourier transforms of $\xi$ and $\zeta$ respectively), 
\beq
\lexp \nt (\k_1) \nt (\k_2) \rexp = \bar n^2 
[(2 \pi)^3 \dD(\k_1 + \k_2)] \, P(k) , 
\eeq
\beq
\lexp \nt (\k_1) \nt (\k_2) \nt (\k_3) \rexp = \bar n^3 
[(2 \pi)^3 \dD(\k_1 + \k_2 + \k_3)] \, B(\k_1, \k_2, \k_3) .
\eeq
The ``momentum conserving'' Dirac $\d$-function, arising from  
homogeneity, ensures that the bispectrum is defined only for 
configurations of wave vectors that form a closed triangle, 
$ \sum \k_i = 0 $.
The reduced three-point amplitude is defined to be 
\beq
Q(\k_1, \k_2, \k_3) = { B(\k_1, \k_2, \k_3) \over 
P(k_1) P(k_2) + P(k_1) P(k_3) + P(k_2) P(k_3) } \label{Q123}
\eeq

The projected Fourier amplitude is 
\beqa
\nt_p (\q) &=& \int d^2x \, n_p(\x) e^{-i \qs \cdot \xs} 
= \int d^2x \, dz \, F(z) \, n(\r) e^{-i \qs \cdot \xs} 
\nonumber \\
&=& \int d^3r \, F(z) \int {d^3 k_z \over (2 \pi)^3} \nt(\k_z) 
e^{i \ks_z \cdot \rs} e^{-i \qs \cdot \xs} 
= \int {dk_z \over 2 \pi} \Ft^\ast (k_z) \, \nt(\q, k_z) , 
\eeqa
the spatial amplitude evaluated at $ \k = \q $ smeared in the projection 
direction over $ \Ft(k_z) = \int dz \, F(z) e^{ik_z z} $.
Note that if $ F $ has typical scale $ \Dstar $, 
then $ \Ft $ has extent $ 1/\Dstar $.

Moments of the projected amplitudes then give the 
corresponding projected spectra.
Normalized by $ \bar n_p $ these are 
\beq
P_p(q ) =  {1\over [\int dz \, F(z)]^2} \int {dk_z \over 2\pi} \, 
{|\Ft(k_z)|^2 } \, P[(q^2 + k_z^2)^{1/2}] \label{Pq} 
\eeq
\beq
B_p(\q_1,\q_2,\q_3) = {1\over [\int dz \, F(z)]^3}
\int {dk_{z1} \over 2\pi} {dk_{z2} \over 2\pi} \, 
\Ft(k_{z1}) \Ft(k_{z2}) \Ft(-k_{z1}-k_{z2}) \, 
B(\q_1,k_{z1};  \q_2, k_{z2}; \q_3, k_{z3}) . \label{Bq}
\eeq
These are the basic results of this section.

The large- and small-$q\Dstar$ limits of these expressions 
are both of interest.
Analogous to the small separation approximation in coordinate space, 
when $ \Dstar $ is large, $ \Ft $ is nonvanishing only for small $k_z$; 
the arguments of $P$ and $B$ are then the transverse $\q$ 
and can be taken outside the integrals.
In this limit the projected spectra are, up to a constant factor, 
the corresponding spatial function evaluated in the transverse plane, 
$ \k \approx (\q,0) $.
By Parseval's theorem and a simple generalization, the integrals 
over $ k_z $ can be written as the same integrals over $z$ 
that appear in equations (\ref{xi_p}) and (\ref{zeta_p}), 
\beq
P_p(q ) \to P(q) \; 
{\int dz \, F^2(z) \over [\int dz \, F(z)]^2} , \label{Pa}
\eeq
\beq
B_p(\q_1,\q_2,\q_3) \to B(\q_1 , \q_2 , \q_3) \; 
{\int dz \, F^3(z) \over [\int dz \, F(z)]^3} , 
\eeq
\beq
Q_p(\q_1,\q_2,\q_3) \to Q(\q_1,\q_2,\q_3) \; 
{[\int dz \, F^3(z)] [\int dz \, F(z)] \over [\int dz \, F^2(z)]^2} 
\label{Qa}
\eeq
($ q \Dstar \gg 1 $).
The opposite limit, large wavelength or small $ \Dstar $, 
corresponds to data confined to a thin sheet or slice taken 
out of three-dimensional space.
In this limit, we have $ \Ft(k_z) \approx \Ft(0) = \int dz \, F(z) $, 
and 
\beq
P_p(q ) \to \int {dk_z \over 2\pi} \, P[(q^2 + k_z^2)^{1/2}]  
\eeq
\beq
B_p(\q_1,\q_2,\q_3) \to  \int {dk_{z1} \over 2\pi} {dk_{z2} \over 2\pi} 
\, B(\q_1,k_{z1}; \q_2,k_{z2}; \q_3,k_{z3}) 
\eeq
($ q \Dstar \to 0 $), where $ k_{z3} = -(k_{z1} + k_{z2}) $.

\section{Numerical Results}

In this section we presents results of numerical integrations of 
equations (\ref{Pq},\ref{Bq}).
For numerical calculations we take $ F(z) = (z/\Dstar) e^{-z/\Dstar} $, 
$ \Ft (k_z) = 1/(1 - i k_z \Dstar)^2 $.
This analytically convenient form mimics the relative 
lack of nearby galaxies because of geometry and the rapid fall 
of the luminosity function above $ L^\ast $.
Because integration is a smoothing operation, we expect the 
general nature of the results to be insensitive to the precise 
behavior of $ F $.
For this $F$, integrals appearing above can be done analytically, 
\beq
I_n = \int dz \, F^n(z) = n! \, (\Dstar/n)^{n+1} , 
\eeq
and so from \eq{Qa} we have for large $ q\Dstar $ 
\beq
{Q_p \over Q} = {I_1 I_3 \over I_2^2} = {32 \over 27} .
\eeq
For our numerical study we use the perturbation theory 
form for the shape dependence of $B$, 
\beq
B(\k_1, \k_2, \k_3) = Q(\k_1,\k_2) P(k_1) P(k_2) 
+ Q(\k_1,\k_3) P(k_1) P(k_3) + Q(\k_2,\k_3) P(k_2) P(k_3) 
\eeq
where 
\beq
Q(\k_i, \k_j) = {10 \over 7} + (\hat{\k}_i \cdot \hat{\k}_j)
\left( {k_i \over k_j} + {k_j \over k_i} \right) 
+ {4 \over 7} (\hat{\k}_i \cdot \hat{\k}_j)^2 
\eeq
(Fry 1984).

Figure~\ref{figP_p} shows the projected $ P_p (q) $ (eq.~[\ref{Pq}]) 
for $ n = +1 $, 0, $-1$, and $-2$, evaluated for $ \Dstar = 1 $.
Results scale with $ \Dstar \neq 1 $ as 
$ P_p(q ; \Dstar) = \Dstar^{-(n+1)} P_p(q\Dstar) $.
Dashed lines show the large-$q$ result (eq.~[\ref{Pa}]), 
$ P_p \sim q^n $.
There is a change of characteristic behavior near $ q = 1 $.
The behavior at small $q$, corresponding to scales much larger 
than the depth of the projection, is that of data on a thin 
two-dimensional plane sheet or slice embedded in three dimensions.

Figure~\ref{figQ_p} shows the dependence of the projected three-point 
amplitude $ Q_p(\theta) $ on configuration shape for triangles 
with sides $ q_1 = q $, $ q_2 = q/2 $, separated by angle $ \theta $.
Curves are shown for $ q \Dstar = 0.001 $, 0.01, 0.1, 
1, 10, 100, and 1000, as indicated in the caption.
The four windows show results for power spectrum indices 
$ n = +1 $, 0, $-1$, and $-2$.
In some cases, the curves for small $ q\Dstar $ overlap and are 
indistinguishable, as is also always the case for large $ q \Dstar $.
In the large-$q\Dstar$ limit the projected result is the result 
in space multiplied by an overall factor, here $ 32/27 $.

Projection changes the shape of of $Q$, making the curves 
in Figure~\ref{figQ_p} flatter at shallow depth, or at 
large wavelength, just as would a bias with $ b > 1 $.
Misconstruing the flattening as an effect of bias gives 
the parameters $ b = b_1 $ and $ b_2/b^2 $ listed in Table 1.
The rms difference between the projected $Q$ and the fit $Q$, 
$ \D^2 = \int d\theta \, (Q_p - Q_b)^2 / \pi $, is small, 
at most $ 0.0036 $, for $ n =-2 $ and $ q\Dstar = 1 $ 
and often much smaller.
The difference between the fit value of $b$ and 1 is of order 
$ 1/q\Dstar $ when $ q\Dstar $ is large.

\section{Angular Projection}

\subsection{Full Sky Coverage}

Typically, observations without distance information give galaxy 
positions not in Cartesian coordinates but in angle on the the sky; 
so we next consider the effects of projection onto the sphere.
In spherical coordinates, a position in space is specified by 
distance $r$ from the origin and angular coordinates $ \theta $,  
$ \phi $, also denoted variously by solid angle $ \Omega $ and 
unit vector $ \nhat $.
Summed over $r$, the projected angular density on the sphere is 
\beq
\N(\theta,\phi) = \int r^2 dr \, F(r) \, n(r,\theta,\phi) , 
\eeq
with average value $ \bar \N = \bar n \int r^2 dr \, F(r) $.

The natural transform on the sky is by spherical harmonics.
For comparison with later results, we compute first the 
full spherical harmonic transform.
First, we compute the amplitude 
\beq
\Nt_{\ell m} = \int d\Omega \, \N(\Omega) \,  Y_{\ell m}^\ast .
\eeq
We wish to to relate moments of this amplitude to the spatial 
power spectrum.
The first step is to express $ \Nt_{\ell m} $ in terms of 
the Fourier amplitude $ \nt(\k) $ defined in equation (\ref{FT}).
Useful relations towards this end include the expansion of a 
plane wave in spherical coordinates, 
\beq
e^{ikz} = \sum_\ell i^\ell (2\ell+1) j_\ell(kr) P_\ell(\cos \theta) 
\eeq
and the addition theorem for spherical harmonics, 
\beq
4 \pi \sum_m Y_{\ell m}(\nhat_1) Y_{\ell m}^\ast(\nhat_2) 
= (2 \ell + 1) \, P_\ell(\nhat_1 \cdot \nhat_2) , 
\eeq
where $ P_\ell $ is the Legendre polynomial of order $\ell$
(Jackson 1975).
From these, we obtain the transform amplitude 
\beq
\Nt_{\ell m} = \int {d^3 k \over (2 \pi)^3} \nt(\k) \, \Ft_\ell(k) 
\, 4 \pi i^\ell \, Y_{\ell m}^\ast(\hat{\k}) , \label{ntlm}
\eeq
where $ \Ft_\ell(k) = \int r^2 dr \, F(r) \, j_\ell(kr) $ is real.

Only $ \Nt_{00} $ has a nonvanishing expectation, 
$ \lexp \Nt_{00} \rexp = \sqrt{4\pi} \bar n \Ft_0(0) $.
The second moment 
$ \lexp \Nt_{\ell_1 m_1} \Nt_{\ell_2 m_2} \rexp $
vanishes unless $ \ell_1 = \ell_2 $, $ m_1 = -m_2 $.
Normalized by $ \lexp \Nt_{00} \rexp $ we have 
\beq
\P_\ell ={\lexp | \Nt_{\ell, m} |^2 \rexp 
\over \lexp \Nt_{00} \rexp^2 } = \int {d^3 k \over (2 \pi)^3} 
{ \Ft_\ell^2(k) \over \Ft_0^2(0) } \, P(k) , \label{P_ell}
\eeq
independent of $m$.

When $ \ell $ is large, the sequence of functions $ \Ft_\ell (k) $ 
approaches a scaling function, 
$ \Ft_\ell (k) \sim \ell^{3/2} \Ft_s (k\Dstar / \ell) $, 
where $ \Ft_s(y) $ is peaked around $ y \sim 1 $ and has 
width $ \D y \sim 1 $, as shown in Figure \ref{figFs}.
With a rapidly decreasing $ F(r) $, this can be understood as 
the main contribution to $ \Ft_\ell (k) $ arising from the first 
peak of $ j_\ell (x) $, which appears at $ k\Dstar \approx \ell $.
With guidance from the small angle limit below, we find 
$ \Ft_s = (\pi/2)^{1/2} y^{-3} e^{-1/y} $, 
which agrees with the solid line so well as to be 
indistinguishable in the figure.
Thus, for large $ \ell $ we have the asymptotic expression 
\beq
\P_\ell \to P_s (\ell) = \int {d^3 y \over (2 \pi)^3} 
{\Ft_s^2(y) \over \Ft_0^2(0)} \, P(\ell y / \Dstar) . 
\label{Ps}
\eeq 
For power law $ P(k) \sim k^n $, equation (\ref{Ps}) 
gives $ P_s(\ell) \sim P(\ell/\Dstar) \sim (\ell/\Dstar)^n$.
Results for $ \P_\ell $ for power law $ P(k) $ are plotted in 
Figure~\ref{figP_l}; the scaling result $ P_s(\ell) $ 
is plotted as dashed lines.
The behavior is similar to that seen in Figure~\ref{figP_p}, 
approaching the scaling result for large $\ell$.

The third moment of the amplitude in \eq{ntlm} is
\beqa
\lexp \Nt_{\ell_1 m_1} \Nt_{\ell_2 m_2} \Nt_{\ell_3 m_3} \rexp &=& \int 
{d^3k_1 \over (2\pi)^3} {d^3k_2 \over (2\pi)^3} {d^3k_3 \over (2\pi)^3} 
\, \Ft_{\ell_1}(k_1) \Ft_{\ell_2}(k_2) \Ft_{\ell_3}(k_3) \, 
(4\pi \bar n)^3  \, i^{\ell_1+\ell_2+\ell_3} \nonumber \\
&& \qquad \times 
[(2\pi)^3 \dD(\k_1 + \k_2 + \k_3)] \, B_{123} \,  
Y_{\ell_1 m_1}^\ast(\hat{\k}_1) 
Y_{\ell_2 m_2}^\ast(\hat{\k}_2) Y_{\ell_3 m_3}^\ast(\hat{\k}_3) .
\eeqa
Indices $ \ell_1 $, $ \ell_2 $, and $ \ell_3 $ specify the overall 
frequency of angular oscillation of $ Y_{\ell m} $, 
while $ m_1 $, $ m_2 $, and $ m_3 $ specify the orientation of the 
oscillation with respect to a chosen polar axis.
Because the universe is believed to be statistically isotropic, 
we are interested in the part of this expression that is invariant 
under rotations.
This can be expressed using properties of the Wigner 3-$j$ coefficients 
(see Edmonds 1957), related to Clebsch-Gordan coefficients by  
\beq
\left( \matrix{ j_1 & j_2 & j_3 \cr m_1 & m_2 & m_3 \cr } \right) = 
{(-1)^{j_1 - j_2 + m_3} \over (2 j_3 + 1)^{1/2}} 
\lexp j_1 \, m_1 \,j_2 \,m_2 \cbar j_3 \, -\!m_3 \rexp .
\eeq
Under a rotation $ D(\omega) $ with Euler angles 
$ \omega = (\alpha,\beta,\gamma) $, states with the same $j$ 
but different values of $m$ mix, 
\beq
D(\omega) \cbar j \, m \rexp = 
\sum_{m'} \DD_{mm'}^{(j)}\cbar j \, m' \rexp 
\eeq
where the $\DD$ are matrix elements of $ D(\omega) $, 
\beq
\DD_{mm'}^{(j)} = \lexp j\, m' \cbar D(\omega) \cbar j \, m \rexp , 
\eeq
with 
\beq
\DD_{m0}^{(\ell)}(\omega) = (-1)^m \left( {4\pi \over 2\ell+1} 
\right)^{1/2} Y_{\ell m}(\beta,\alpha) .
\eeq
In terms of the rotation matrix elements, the effect of a rotation 
on the 3-$j$ coefficient is 
\beq
\sum_{m'_1,m'_2,m'_3} 
\DD_{m'_1 m_1}^{(j_1)} \DD_{m'_2 m_2}^{(j_2)} \DD_{m'_3 m_3}^{(j_3)} 
\left( \matrix{ j_1 & j_2 & j_3 \cr m'_1 & m'_2 & m'_3 \cr } \right) = 
\left( \matrix{ j_1 & j_2 & j_3 \cr m_1 & m_2 & m_3 \cr } \right) ; 
\eeq
if we take $ m_1 = m_2 = m_3 = 0 $, we then have 
\beq
\sum_{m_1,m_2,m_3} \left( \matrix{ \ell_1 & \ell_2 & \ell_3 \cr 
m_1 & m_2 & m_3 \cr } \right) Y_{\ell_1 m_1} Y_{\ell_2 m_2} Y_{\ell_3 m_3} 
= \left[ {(2\ell_1+1) (2\ell_2+1) (2\ell_3+1)\over (4\pi)^3 } \right]^{1/2}
\left( \matrix{ \ell_1 & \ell_2 & \ell_3 \cr 0 & 0 & 0 \cr } \right) .
\eeq
Nonvanishing contributions to the sum have $ m_1 + m_2 + m_3 = 0 $; 
the $ \ell_i $ must also obey triangle inequalities, 
$ | \ell_1 - \ell_2 | \leq \ell_3 \leq \ell_1 + \ell_2 $, etc.; 
and the sum $ \ell_1 + \ell_2 + \ell_3 $ must also be even.
Removing the dependence on $m$ leaves a result independent of 
orientation of the polar axis.
This leads us to consider the averaged angular bispectrum, 
\beqa
\B_{\ell_1 \ell_2 \ell_3} &=& \sum_{m_1,m_2,m_3} \left (
\matrix{ \ell_1 & \ell_2 & \ell_3 \cr m_1 & m_2 & m_3 \cr} \right) 
{\lexp \Nt_{\ell_1 m_1} \Nt_{\ell_2 m_2} \Nt_{\ell_3 m_3} \rexp 
\over \lexp \Nt_{00} \rexp ^3 } \nonumber \\ 
&=& I_{\ell_1\ell_2\ell_3} \, \int 
{d^3k_1 \over (2\pi)^3} {d^3k_2 \over (2\pi)^3} 
{\Ft_{\ell_1}(k_1) \over \Ft_0(0)}{\Ft_{\ell_2}(k_2) \over \Ft_0(0)}
{\Ft_{\ell_3}(k_3) \over \Ft_0(0)} \, B(\k_1,\k_2,\k_3) , 
\eeqa
where $ \k_3 = -\k_1 - \k_2 $ and the coefficient 
$ I_{\ell_1\ell_2\ell_3} $ is 
\beq
I_{\ell_1\ell_2\ell_3} = 
\left[ {(2\ell_1+1) (2\ell_2+1) (2\ell_3+1)} \right]^{1/2}
\left( \matrix{ \ell_1 & \ell_2 & \ell_3 \cr 0 & 0 & 0 \cr } \right) .
\eeq
This angular averaging was introduced by Spergel \& Goldberg (1998) 
for the microwave background.
This is formally very similar to the Cartesian result, \eq{Bq}.
For the behavior at large $\ell$  we look at partial sky coverage 
in the small angle approximation in the next section.

\subsection{Partial Sky Coverage}

In practice, there is no galaxy or cluster catalog with complete sky 
coverage, and in deep catalogs interesting scales cover only small angles.
It is far more efficient computationally to treat a small square patch 
of sky as flat and use a Fourier transform, or especially FFT, 
rather than the spherical harmonic transform (Fry \& Seldner 1982).
Our final goal is to determine just what the result is 
of such a procedure.

The Fourier transform in ``flat'' angular coordinates $ \bth $ is 
\beq
\nt_p(\bkp) = \int_\Lambda d^2\theta \, n_p(\bth) \, 
e^{-i \bkps \cdot \bths} . 
\eeq
where $ \bth=(\theta_1,\theta_2) = (\theta\cos\phi,\theta\sin\phi) $ 
runs over a small patch of sky, say a square of side $ \Lambda $, 
and $ \bkp $ is a Fourier frequency conjugate to $ \bth $.
The projected amplitude $ \nt_p(\bkp) $ in terms of 
$ \nt(\k) $ is then in full 
\beq
\nt_p(\bkp) = \int_\Lambda d^2\theta \, e^{-i \bkps \cdot \bths} 
\int {d^3 k \over (2 \pi)^3} \, \Ft_\ell(k) \, \nt(\k) 
\sum_{\ell,m}  4 \pi \, i^\ell Y_{\ell m} (\nhat) 
Y_{\ell m}^\ast (\hat{\k}) , \label{nkappa}
\eeq
For the full sky transform, the integrals over angles involved 
orthogonal functions, leading to the simple result in \eq{ntlm}.
The projected second moment of $ \nt_p $ here is 
\beq
\lexp \nt_p(\bkp_1) \nt_p(\bkp_2) \rexp = \bar n_p^2 
\sum_{\ell=0}^\infty  \D_\ell(\bkp_1,\bkp_2) \, \P_\ell 
\eeq 
where $ \P_\ell $ is as in \eq{P_ell} and 
\beq
\D_\ell(\bkp_1,\bkp_2) = (2 \ell + 1) 
\int_\Lambda d^2 \theta_1 d^2 \theta_2 \, 
e^{-i \bkps_1 \cdot \bths_1} e^{-i \bkps_2 \cdot \bths_2} 
P_\ell(\nhat_1 \cdot \nhat_2) .  \label{D_l}
\eeq 
This integral behaves as an effective $\d$-function, 
peaked for $ \bkp_1 + \bkp_2 = 0 $ and at $ \ell = |\kappa| $, 
but smeared in $ \kappa $ and $ \ell $ over roughly  
$ \pm (2\pi / \Lambda) $, or $ \pm 1 $ wave per box.
This is shown in Figure \ref{figDelta} for $ \bkp = (8,0) $, 
$ (7,4) $, and $ (5,6) $, in units of $ 2\pi/\Lambda = 18.947 $, 
each with approximately the same magnitude, $ \kappa = 8 $, 8.062, 
and 7.810, respectively, but different orientations.
Both the location and width of the peaks behave as expected.

Useful expressions can be obtained when $ \Lambda $ is small, 
so that $ \r = (r \theta \cos \phi, r \theta \sin \phi , r) $.
Integrals over angles such as 
\beq
\int_{-\Lambda/2}^{+\Lambda/2} d\theta_x \, e^{i(k_x r - \kappa_x)} 
= {\sin[\Lambda(k_x r - \kappa_x)/2] \over (k_x r - \kappa_x)/2 } 
\eeq
are effectively $\d$-functions, with resolution in $ \kappa $ 
of $ \pm 2 \pi /\Lambda $.
With this, the projected amplitude becomes 
\beq
\nt_p(\bkp) 
= \int {dk \over 2\pi} \int dr \, F(r) \, e^{ikr} \, \nt(\bkp/r,k)  
= \int {d\lambda \over 2\pi} \int {du \over u^2} \, F(1/u) \, 
e^{i\kappa\lambda} \, \nt(\bkp u, \kappa u \lambda) , 
\eeq
where $ r = 1/u $ and $ k = \lambda \kappa u $.
Then we have for the second moment 
\beq
\lexp \nt_p(\bkp_1) \nt_p^\ast(\bkp_2) \rexp = 
\bar n^2 (2 \pi)^2 \dD(\phi_1-\phi_2) {1 \over \kappa_2} 
\int {d \lambda \over 2 \pi} \, e^{i\lambda\Dk}  
\int {du_1 \over u_1^2 u_2^2} \, F(1/u_1) F(1/u_2) \, 
P[\kappa_1 u_1 (1+\lambda^2)^{1/2}] ,  \label{n_p^2}
\eeq
where the $\d$-function in the second moment of $ \nt(\k) $ 
require that $ \bkp_1 $ and $ \bkp_2 $ are in opposite directions 
and thus $ u_2 \kappa_2 = u_1 \kappa_1 $.
For power law $ P(k) \sim k^n $ the power spectrum factors 
to $ (1+\lambda^2)^{n/2} P(\kappa_1 u_1) $.
The integral over $ \lambda $ can be expressed in terms 
of modified Bessel functions of order $ \pm (n+1)/2 $ 
for non-integer $n$, 
while for negative integer $n$, 
\beq
\int {d\lambda \over 2 \pi} \, e^{i\lambda\Dk} \, 
(1+\lambda^2)^{n/2} = \cases { 
\dD(\Dk) & $ n =  0 $, \cr
\noalign{\vskip 2pt}
{1 \over \pi} K_0(|\Dk|) & $ n = -1 $, \cr
\noalign{\vskip 2pt}
{1 \over 2} e^{-|\Dk|} & $ n = -2 $, \cr
\noalign{\vskip 2pt}
{1 \over \pi} |\Dk| \, K_1 (|\Dk|) & $ n = -3 $. \cr }
\label{lambda} .
\eeq
All of these are appreciable only for $ |\Dk| \la 1 $, fall off 
exponentially for large $ \Dk $, and integrate to 1; so 
for any $n$ the integral is once again effectively a $\d$-function 
of $\Dk$, to an accuracy $ \pm 1 $, where for small $ \Lambda $, 
$ \kappa $ is large.
Requiring $ \kappa_2 = \kappa_1 $ implies $ u_2 = u_1 $ also, 
and thus the second moment of the projected Fourier amplitude 
in \eq{n_p^2} becomes 
\beq
\lexp \nt_p(\bkp_1) \nt_p(\bkp_2) \rexp = 
\bar n_p^2 \, [(2 \pi)^2 \dD(\bkp_1+\bkp_2)] \, P_p(\kappa) , 
\eeq
where 
\beq
P_p(\kappa) = {1 \over [\int F(1/u) du/u^4]^2} 
\int {du \over u^4} \, F^2 (1/u) \, P(\kappa u) . \label{Pkappa}
\eeq
Although this expression was obtained for a power law $ P(k) $, 
to the extent that the integral over $ \lambda$ in \eq{lambda} 
is effectively a $\d$-function, 
this ought to apply for any well-behaved power spectrum.
This equation is the same Kaiser's (1992) equation (A.2), obtained 
here by a different path, and is equivalent to the scaling result in 
equation (\ref{Ps}) for $P_s (\ell)$ at large $ \ell $. 
For $ F(r) = \exp(-r/\Dstar) $, this gives us the form of the 
scaling function, $ \Ft_s (y) = (\pi/2)^{1/2} y^{-3} e^{-1/y} $.

For the third moment, we have 
\beqa
\lexp \nt_p(\bkp_1) \nt_p(\bkp_2) \nt_p(\bkp_3) \rexp &=& 
\bar n^3 \int {dk_1 \over 2\pi}{dk_2 \over 2\pi}{dk_3 \over 2\pi}
\int {du_1 \over u_1^2}{du_2 \over u_2^2}{du_3 \over u_3^2} 
F(1/u_1) F(1/u_2) F(1/u_3) \nonumber \\
&& \quad \times [(2 \pi)^3 \dD(\k_1+\k_2+\k_3)] \, 
B(u\bkp_1, k_1; u\bkp_2, k_2; u\bkp_3,k_3 ) , 
\eeqa
where as before $ u = 1/r $.
One component of the momentum-conserving $\d$-function constrains 
$ k_1 + k_2 + k_3 = 0 $; the two others $ \sum u_i\bkp_i = 0 $.
One of these last, in any direction not perpendicular to $ \bkp_1 $, 
can be taken to give the value of $ u_1 $, and the second component 
to give $ u_2 $, in terms of $ u_3 $ and the $ \bkp_i $, leaving 
\beqa
\lexp \nt_p(\bkp_1) \nt_p(\bkp_2) \nt_p(\bkp_3) \rexp &=& 
\bar n^3 \int {d k_1 \over 2\pi} e^{i k_1 \Dk_{13}}  
\int {d k_3 \over 2\pi} e^{i k_3 \Dk_{23}} 
\int {du_1 \over u_1^2 u_2^2 u_3^2} \nonumber \\ 
&& \quad \times \, F(1 / u_1) F(1 / u_2)  F(1 / u_3) 
\, B(u_1 \bkp_1 , u_2 \bkp_2 , u_3 \bkp_3 ) 
\eeqa
As was found for the second moment, the $k$-integrals are 
effectively $\d$-functions, enforcing constraints that require  
the triangle in the $ \bkp_i $ to be closed; when these are 
satisfied all three $u_i$ are equal.
Thus, we have finally 
\beq
\lexp \nt_p(\bkp_1) \nt_p(\bkp_2) \nt_p(\bkp_3) \rexp = 
\bar n_p^3 [(2 \pi)^2 \dD(\bkp_1+\bkp_2+\bkp_3)] \, 
B_p(\bkp_1, \bkp_2, \bkp_3 ) , 
\eeq
\beq
B_p(\bkp_1, \bkp_2, \bkp_3) =  {1 \over [\int du \, F(1/u) /u^4]^3}
\int {du \over u^4} \, F^3(1/u) \, B(u\bkp_1, u\bkp_2, u\bkp_3 ) .
\label{Bkappa}
\eeq
This is what we have been aiming for, an expression for the 
projected bispectrum in terms of the full bispectrum in 
space and the selection function $F$.
For power law $ P(k) \sim k^n $, $ B (u \bkp_i) = u^{2n} B(\bkp_i) $, 
and so the projected bispectrum has the same shape dependence 
as the bispectrum in space, with an overall multiplicative factor 
\beq
Q_p = {[\int du \, F^3(1/u) \, u^{2n-4}] [\int du \, F(1/u) \, u^{-4}] 
\over [\int du \, F^2(1/u) \, u^{n-4}]^2} Q_{123} . \label{QGP}
\eeq
Written in terms of $ r = 1/u $, this is the same projection factor 
obtained by Groth \& Peebles (1977) for $Q$ in space and 
by Bernardeau (1995) and Gazta\~{n}aga \& Bernardeau (1998) 
for the skewness $S_3$.

As is apparent in equations (\ref{Pkappa}) and (\ref{Bkappa}), 
at fixed angles we are probing clustering on physical scales 
$ \Delta r = r \Delta \theta $; and as $r$ ranges over  
a broad selection function this can be a substantial variation.
For a power spectrum that is not a simple power law, the effect 
of projection depends on how much the effective index 
$ n_\eff = d \log P / d \log k $ varies over the extent 
of the selection function.
An example is shown in Figures \ref{figPkappa} and \ref{figQkappa} 
for the cold dark matter (CDM) power spectrum, as parametrized by 
Bardeen et al. (1986), for $ \Omega = 0.3 $, $ h = 0.7 $.
The CDM spectrum has $ n = +1 $ as $ k \to 0 $ 
and $ n \to -3 $ as $ k \to \infty $, and the projected 
spectrum has these limits also.
Where the effective index is almost constant, the projected power is 
the same as would be obtained for a power-law spectrum with the same $n$.
However, for values of $ \kappa $ near the peak the effective index 
varies enough to give a visibly different result.
For the bispectrum, the effects of projection include a change 
in the amplitude of the variation with angle, the effect of a 
linear bias parameter $b \neq 1 $; or as an overall shift in value, 
similar to a quadratic bias parameter $b_2 \neq 0 $; 
or a change in the dependence on configuration that is not 
well fit by a local bias.
Over the range of scales of the Lick catalog (Fry 1994), 
which corresponds to $ \kappa = 25 $--150, 
the apparent bias parameters in Table~2 differ from 1 by 
about 10\%  to only 2\%  at the scales most heavily weighted.

\section{Discussion} 

In this paper we have investigated the effects of projection on 
the power spectrum and bispectrum in a variety of geometries.
In general, projection is a smoothing operation, smearing 
out the dependence of the power spectrum and bispectrum on scale 
and the dependence of the bispectrum on configuration shape.

The degree of smoothing depends on the depth of the projection.
For projection along a Cartesian direction we observe a range 
of behavior as this depth varies from very shallow to very deep.
On scales small compared to the depth or wavenumbers $q$ 
such that $ q\Dstar $ is large, except for an overall numerical 
factor projection has minimal effect; 
but the result of smoothing becomes increasingly visible as 
$q$ decreases, with a qualitative change in behavior at 
$ q\Dstar \approx 1 $.
For power law $ P(k) \sim k^n $, the projected power spectrum is also 
$ P_p(q) \sim q^n $ with the same value of $n$ for large $ q\Dstar $ 
but the slope becomes noticeably shallower for small $ q \Dstar $.
For the bispectrum we can quantify the result by the 
apparent bias factors given in Table~1.
For large $ q\Dstar $ the apparent bias parameters differ from 
$ b = 1 $ and $ b_2 = 0 $ by an amount of order $ 1/ q\Dstar $ 
or smaller.
This can be traced to equations (\ref{Pq}) and (\ref{Bq}), 
where it can be seen that the range of scales contributing has 
spread $ \D k/k \sim 1/q\Dstar$.

The projected full sky angular distribution has power $ \P_\ell $ 
in spherical harmonic amplitudes as given in \eq{P_ell}.
For a power-law power spectrum, the scaling limit in \eq{Ps} gives 
$ \P_\ell \to P_s(\ell) \sim \ell^n $ with the same $n$, while the exact 
result at small $\ell $ departs from this scaling limit in a way similar 
to the behavior of Cartesian result as $ q\Dstar \to 1 $.
Restricting the angular transform to a small portion of the sky of 
size $ \Lambda $ smears the result over values of $ \ell $ over a 
range $ \pm 2 \pi/\Lambda $.
Once again the limiting behavior for large $ \kappa $, where 
$ \kappa $ is a transform variable conjugate to $ \theta $, 
is $ P_p(\kappa) \sim \kappa^n $.

Additional effects enter when the power spectrum is not 
a pure power law,
The integrals in \eq{Pkappa} and \eq{Bkappa}, and the obvious 
generalization for the $n$-point spectrum 
\beq
B_{n,p} (\bkp_1, \dots, \bkp_n) = {1 \over [\int du \, F(1/u) /u^4]^n} 
\int {du \over u^4} F^n(1/u) \, B_n(u\bkp_1, \dots, u\bkp_n ) 
\eeq
($ \sum \bkp_i = 0 $), average the spatial function over the 
range of scales covered by the selection function $ F(r) $.
When the effective index $ n_\eff = d \log P/ d\log k $ has 
little dependence on scale, the power-law results still remain 
approximately valid, but when $ n_\eff $ is changing appreciably 
over the scales sampled by $F$ the projected functions can 
vary greatly, manifesting as apparent values $b$ and $b_2$.

From what we have learned above, to avoid introducing artifacts 
of projection in analyses of data we should limit consideration to 
scales where $ \kappa \gg 1 $, and for angular transform 
on a domain of side $ \Lambda $ to $ \kappa \Lambda > 1 $ as well.
These restrictions appear to be satisfied in the Lick analysis (Fry 1994), 
for which $ \kappa = 25 $--152 and $ \kappa \Lambda = 31 $--188.
This would appear to suggest that the weak shape dependence of 
the Lick bispectrum (Fry \& Seldner 1982; Fry 1994) cannot be entirely 
ascribed to projection (but could be from nonlinear evolution, 
Scoccimarro et al. 1998).

Planned future microwave background experiments will eventually 
lead to precise measurements of cosmological parameters such as 
the Hubble constant $ H_0 $, the fraction of critical density 
$ \Omega_0 $, and the cosmological constant $ \Lambda $, and also 
of properties of the primordial power spectrum such as spectral 
index $n$ and normalization $Q$.
On the scales of galaxy clustering, what we will then be studying will 
be details of the modulation of,primordial power by the dark matter.
In order to do that, we must be able to measure bias parameters.
Study of the galaxy bispectrum appears to be one tool with the 
potential to do this, and projection does not appear to introduce 
unsurmountable difficulties.

\acknowledgments

A portion of this work was done while J.N.F. was 
a visitor at the Aspen Center for Physics.
A similar subject has been addressed recently by 
Buchalter, Kamionkowski, \& Jaffe (1999).
Research supported in part by NASA grant NAG5-2835 at the 
University of Florida.

\begin{deluxetable}{rrrrr}
\tablewidth{5in}
\tablecaption{Cartesian Projection Bias Parameters 
\label{table1}}
\tablehead{ \colhead{$n$} & \colhead{$q\Dstar$} & \colhead{$1/b$} 
& \colhead{$ b_2/b^2 $} & \colhead{$ \D_{\rm rms} $} }
\startdata
$+1$ & 0.01 & $-0.0015 $ & $ 0.3598 $ & $ 8.65 \times 10^{-6} $ \nl
     &  0.1 & $-0.0668 $ & $ 0.4159 $ & $ 4.18 \times 10^{-4} $ \nl
     &    1 & $ 0.2911 $ & $ 0.4363 $ & $ 5.37 \times 10^{-4} $ \nl
     &   10 & $ 0.9858 $ & $ 0.0128 $ & $ 1.14 \times 10^{-4} $ \nl
     &  100 & $ 0.9993 $ & $ 0.0004 $ & $ 1.23 \times 10^{-5} $ \nl
     & 1000 & $ 0.9997 $ & $ 0.0003 $ & $ 1.19 \times 10^{-5} $ \nl 
\noalign{\smallskip}
$0$ & 0.01 & $-0.0023 $ & $ 1.1689 $ & $ 2.83 \times 10^{-4} $ \nl
    &  0.1 & $ 0.0031 $ & $ 1.0335 $ & $ 1.49 \times 10^{-3} $ \nl
    &    1 & $ 0.3605 $ & $ 0.4826 $ & $ 5.50 \times 10^{-4} $ \nl
    &   10 & $ 0.9493 $ & $ 0.0358 $ & $ 3.10 \times 10^{-4} $ \nl
    &  100 & $ 0.9984 $ & $ 0.0013 $ & $ 5.67 \times 10^{-5} $ \nl
    & 1000 & $ 0.9993 $ & $ 0.0009 $ & $ 5.56 \times 10^{-5} $ \nl 
\noalign{\smallskip}
$-1$ & 0.01 & $ 0.0272 $ & $ 1.4424 $ & $ 9.04 \times 10^{-4} $ \nl
     &  0.1 & $ 0.0923 $ & $ 1.0868 $ & $ 1.48 \times 10^{-3} $ \nl
     &    1 & $ 0.4519 $ & $ 0.4341 $ & $ 2.28 \times 10^{-3} $ \nl
     &   10 & $ 0.9511 $ & $ 0.0361 $ & $ 8.49 \times 10^{-4} $ \nl
     &  100 & $ 0.9984 $ & $ 0.0017 $ & $ 1.49 \times 10^{-4} $ \nl
     & 1000 & $ 0.9993 $ & $ 0.0012 $ & $ 1.46 \times 10^{-4} $ \nl 
\noalign{\smallskip}
$-2$ & 0.01 & $ 0.2097 $ & $ 0.9742 $ & $ 5.94 \times 10^{-4} $ \nl
     &  0.1 & $ 0.2521 $ & $ 0.8458 $ & $ 5.17 \times 10^{-4} $ \nl
     &    1 & $ 0.5250 $ & $ 0.3788 $ & $ 3.60 \times 10^{-3} $ \nl
     &   10 & $ 0.9510 $ & $ 0.0372 $ & $ 1.33 \times 10^{-3} $ \nl
     &  100 & $ 0.9984 $ & $ 0.0020 $ & $ 2.34 \times 10^{-4} $ \nl
     & 1000 & $ 0.9993 $ & $ 0.0016 $ & $ 2.29 \times 10^{-4} $ \nl 
\enddata
\end{deluxetable}

\begin{deluxetable}{rrrrr}
\tablewidth{5in}
\tablecaption{Angular Projection Bias Parameters 
\label{table2}}
\tablehead{ \colhead{$\kappa$} & \colhead{$n_\eff$} & \colhead{$1/b$} 
& \colhead{$ b_2/b^2 $} & \colhead{$ \D_{\rm rms} $} }
\startdata
  $10$ & $-0.543$ & $ 1.081 $ & $ 0.183 $ & $ 1.34 \times 10^{-2} $ \nl
  $25$ & $-1.289$ & $ 1.019 $ & $ 0.167 $ & $ 8.17 \times 10^{-3} $ \nl
  $50$ & $-1.733$ & $ 0.993 $ & $ 0.109 $ & $ 3.84 \times 10^{-3} $ \nl
 $100$ & $-2.026$ & $ 0.981 $ & $ 0.068 $ & $ 1.37 \times 10^{-3} $ \nl
 $150$ & $-2.152$ & $ 0.978 $ & $ 0.054 $ & $ 6.62 \times 10^{-4} $ \nl
 $500$ & $-2.405$ & $ 0.977 $ & $ 0.029 $ & $ 3.53 \times 10^{-4} $ \nl
$1000$ & $-2.499$ & $ 0.980 $ & $ 0.022 $ & $ 3.74 \times 10^{-4} $ \nl
\enddata
\end{deluxetable}

\clearpage

\plotone{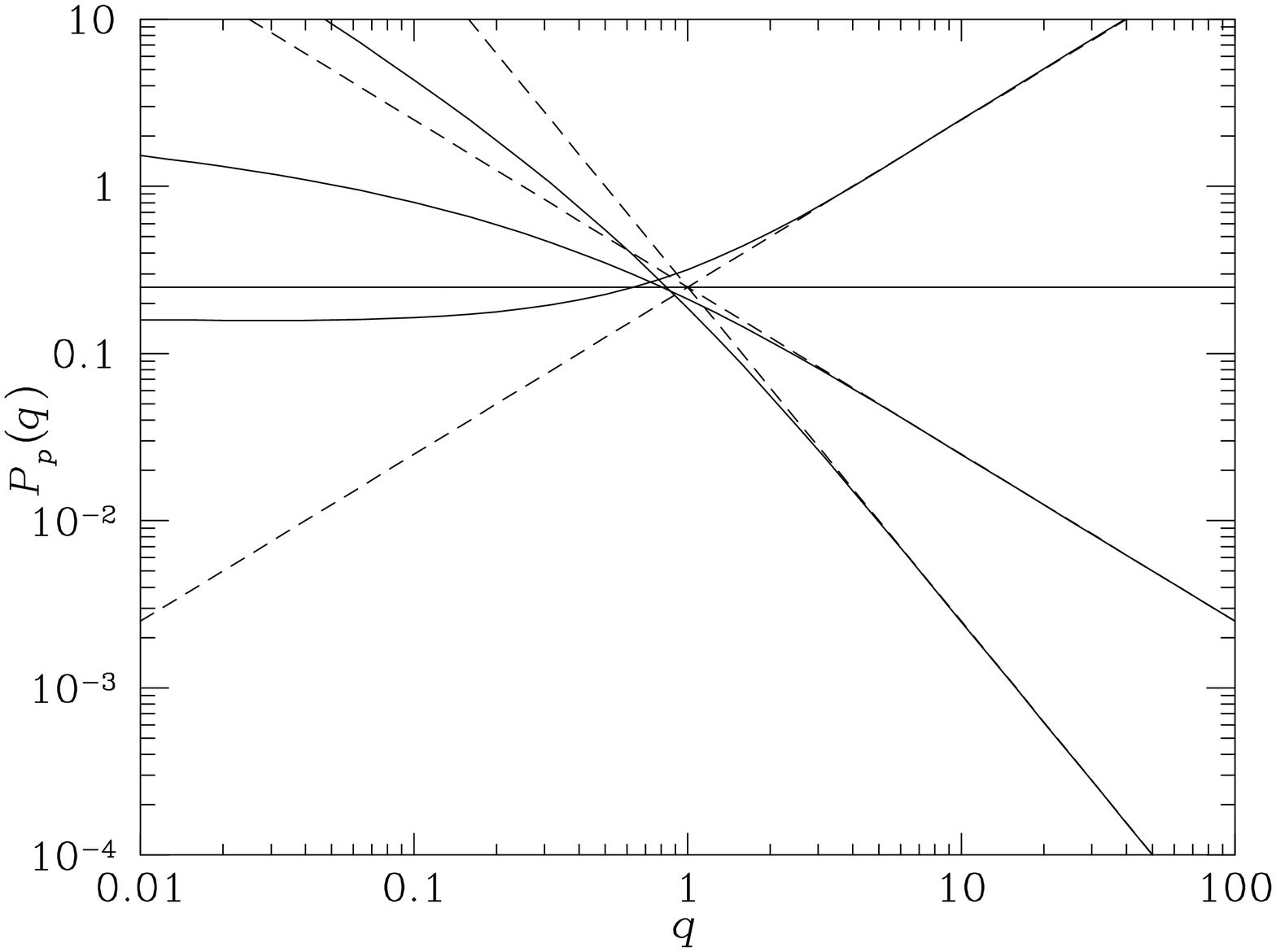}
\figcaption
{Projected $ P_p(q) $ for $ n = +1 $, 0, $-1$, $-2$ 
evaluated for $ \Dstar = 1 $.
Dashed lines show the large-$q$ limit.
\label{figP_p}}

\plotone{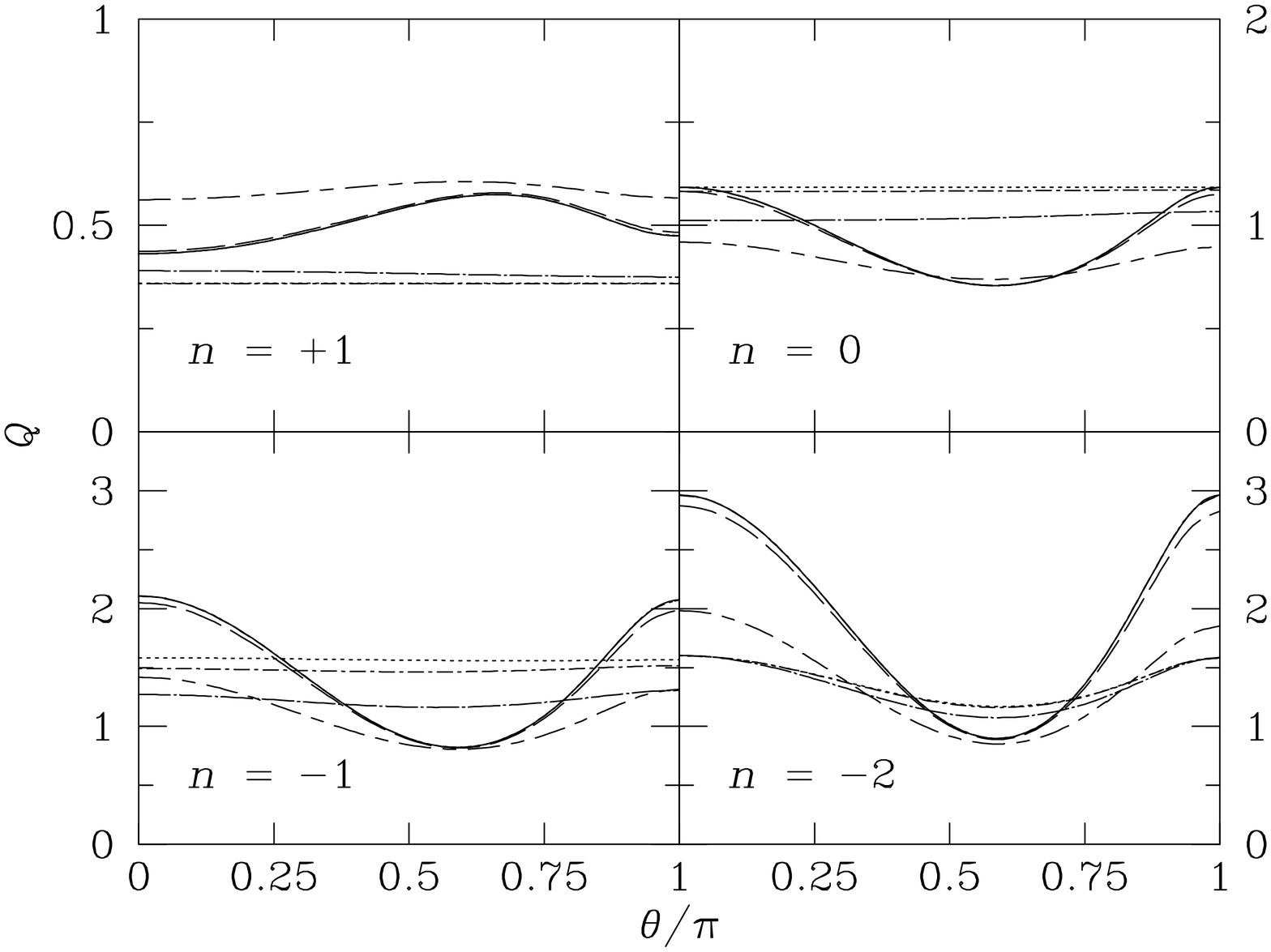}
\figcaption
{Projected $ Q_p(\theta) $ for configurations with $ q_1 = 1 $, 
$ q_2 = 1/2 $, separated by $ \theta $, for power-law power spectra 
$ P \sim k^n $ with $ n = +1 $, $0$, $-1$, $-2$ as labelled.
In each window, the solid curve shows the unprojected result 
scaled by the factor $ 32/27 $; 
the dotted curve shows the projected result for $ \Dstar = 0.001 $; 
the dot-short dash curve $ \Dstar = 0.01 $; 
dot-long dash curve $ \Dstar = 0.1 $; 
short dash-long dash curve $ \Dstar = 1 $; 
long dashed curve $ \Dstar = 10 $.
Results for $ \Dstar = 100 $ and $ \Dstar = 1000 $ 
are indistinguishable from solid curve.
\label{figQ_p}}

\plotone{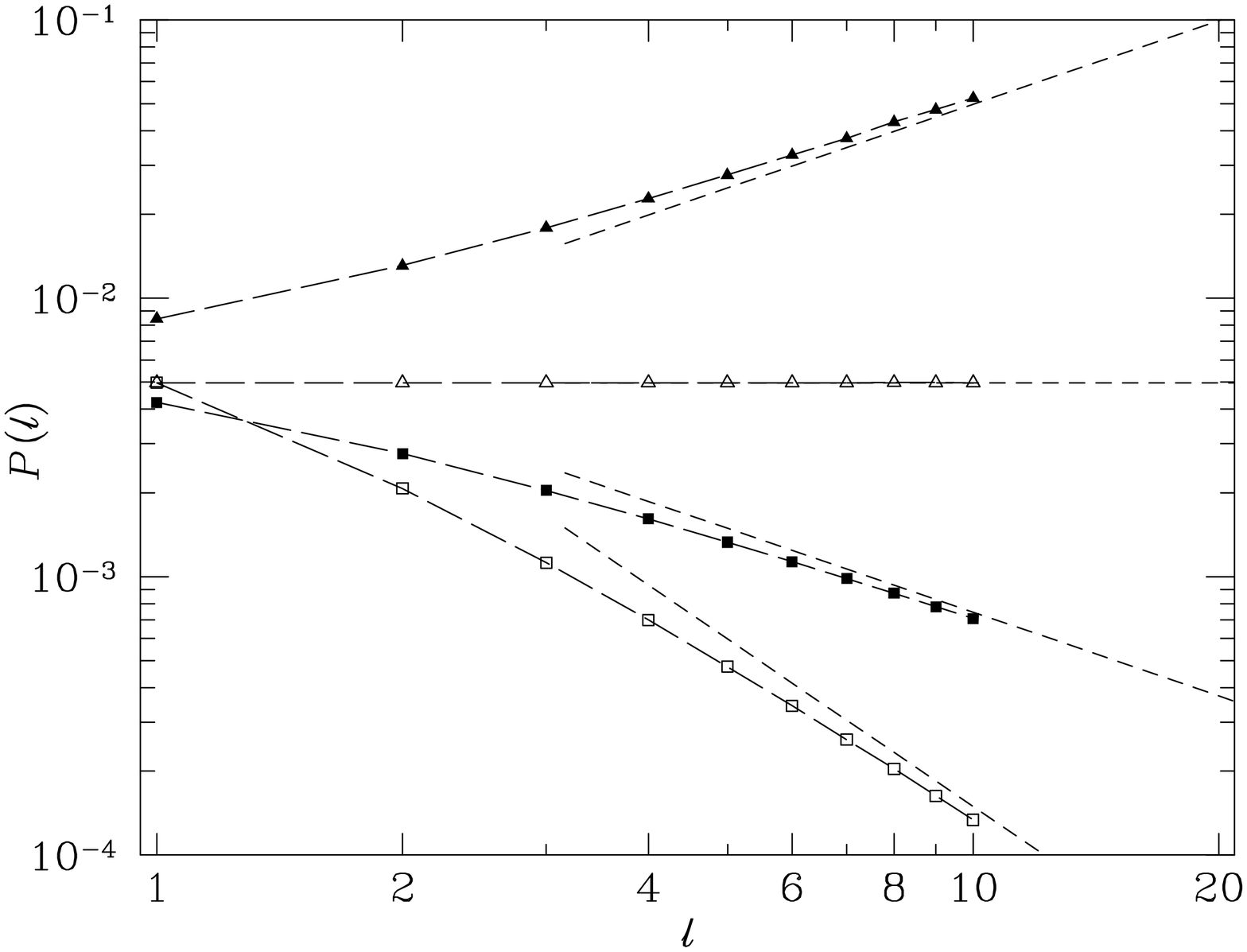}
\figcaption
{Projected angular power $ \P_\ell $ vs.~$\ell$.
Symbols show integrals over spherical Bessel functions 
for $ \ell = 0 $ through 10.
Filled triangles denote $ n = +1 $, open triangles $ n = 0 $, 
filled squares $ n = 0 $, and open squares $ n = -2 $.
Dashed lines show scaling result expected to hold 
for large $\ell$. \label{figP_l}}

\plotone{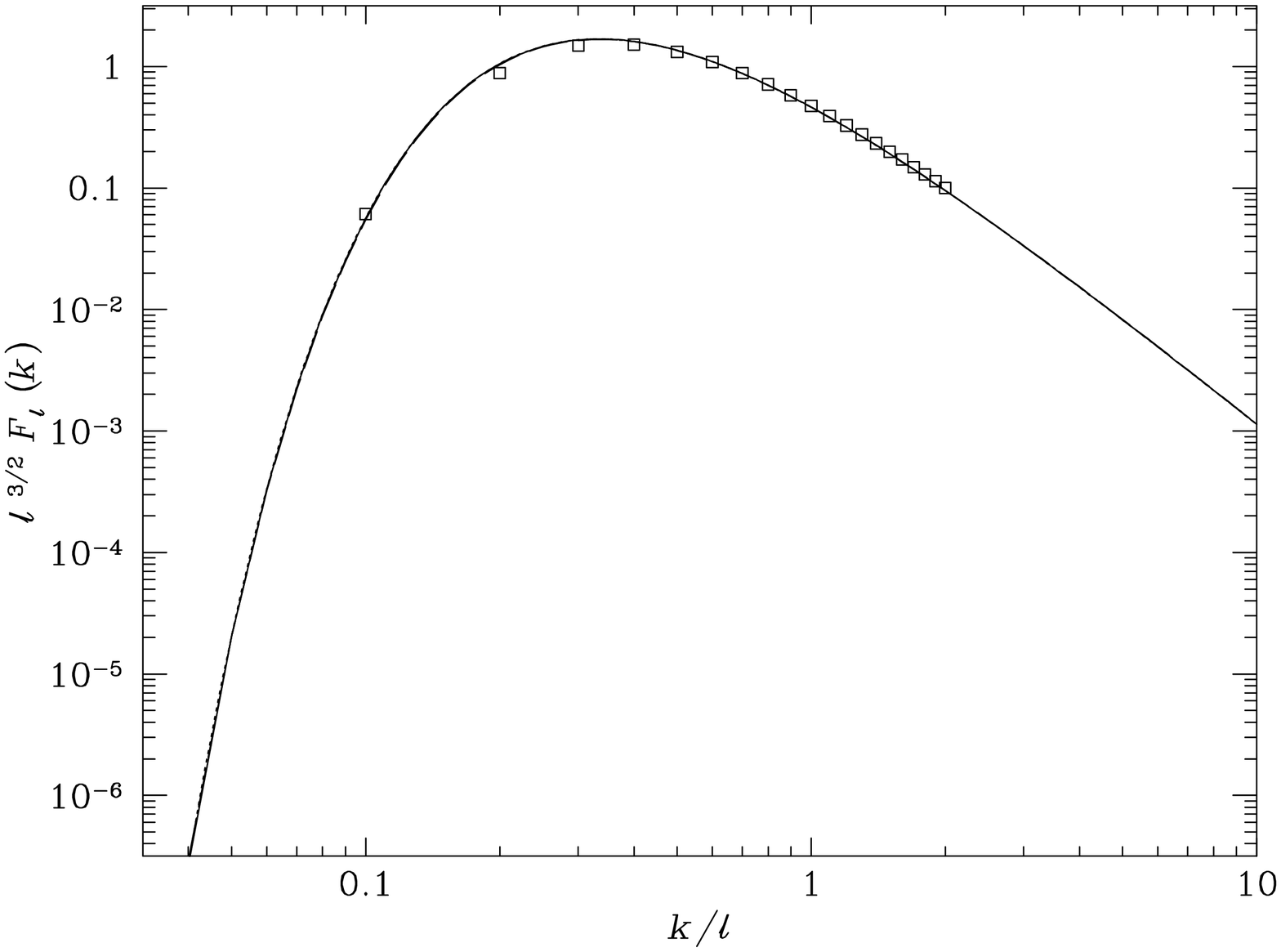}
\figcaption
{Scaled $ \ell^{3/2} \Ft_\ell $ vs. $ k/\ell $.
Symbols show $ \ell = 10 $, dashed curve $ \ell = 100 $, 
solid curve $ \ell = 1000 $.
The two curves are essentially indistinguishable. \label{figFs}}

\plotone{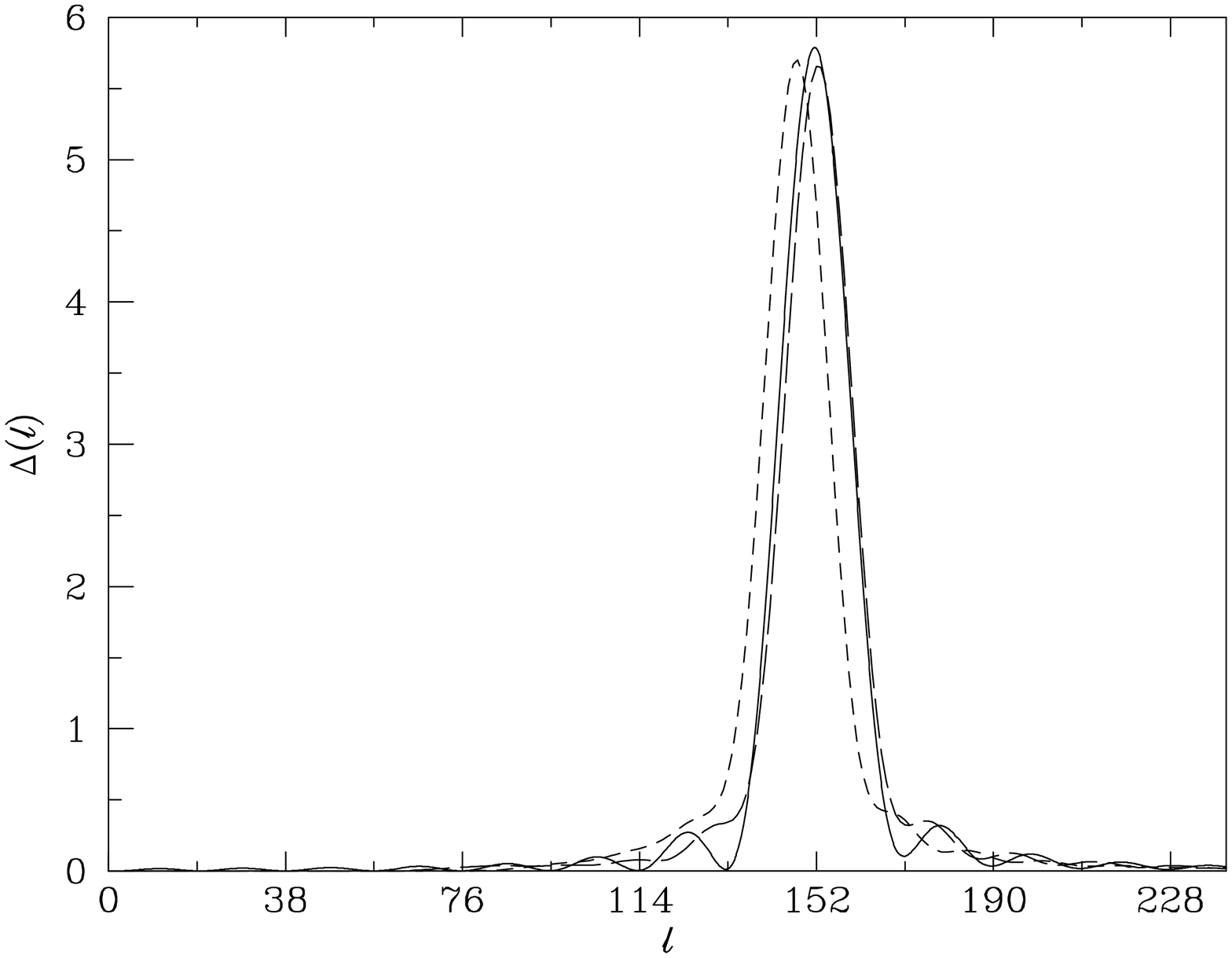}
\figcaption
{Integral $ \D_\ell(\bkp_1,\bkp_2) $ (eq.~[\protect{\ref{D_l}}]) 
plotted vs.~$ \ell $. .
Solid line shows $ \bkp_1=(8,0) $ and $ \bkp_2=(-8,0) $, 
long-dashed line shows $ \bkp_1=(7,4) $ and $ \bkp_2=(-7,-4) $, 
and short-dashed line shows $ \bkp_1=(6,5) $ and $ \bkp_2=(-6,-5) $, 
in units of $ 2\pi/\Lambda $ with $ \Lambda = 19^\circ $.
\label{figDelta}}

\plotone{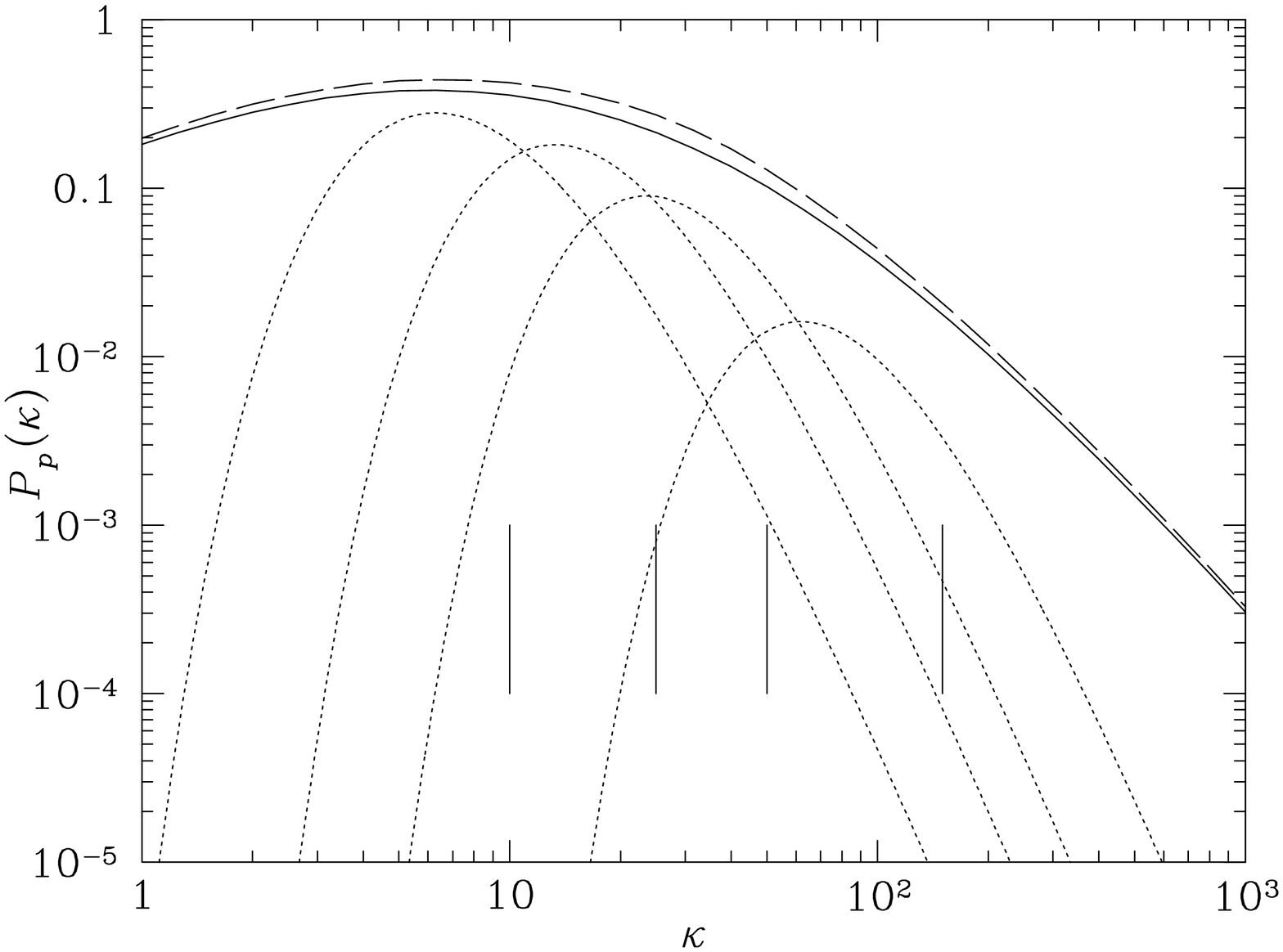}
\figcaption
{Projected $ P_p(\kappa) $ obtained from \protect{\eq{Pkappa}} 
for the CDM power spectrum.
The solid line shows the full integration of the CDM spectrum, 
while the long-dashed line shows what would be obtained for a power 
law spectrum with the effective local index $ n_\eff $.
Vertical lines mark the values $ \kappa = 10 $, 25, 50, and 150 
used in the next figure.
The dotted lines show the spread of scales that contribute to 
the integrand, $ \kappa^3 x^{-3} F^2(-\kappa/x) P(x/\Dstar) $ 
with $ x = u/\kappa\Dstar $, 
for these four values of $ \kappa $.
\label{figPkappa}}

\plotone{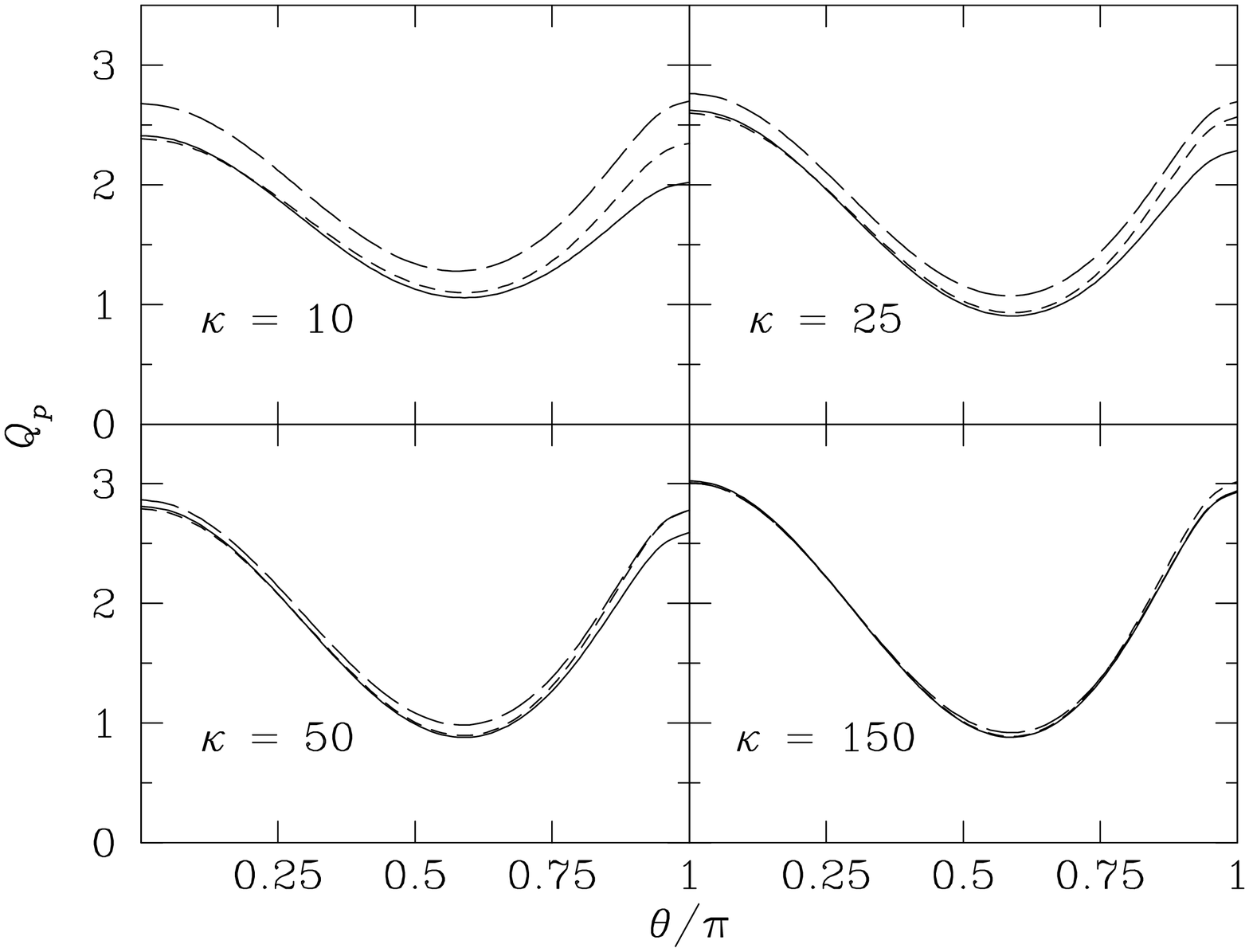}
\figcaption
{Projected $ Q(\theta) $ obtained from \protect{\eq{Bkappa}} 
integrating over the CDM power spectrum for configurations with 
$ \kappa_2 = \frac12 \kappa_1 $ separated by angle $ \theta $ 
for four values of $ \kappa = \kappa_1 $, as indicated.
The long-dashed line shows the actual integrated projection, 
the short-dashed line shows projection of a power-law spectrum with 
index $ n_\eff $ evaluated at $ \kappa_1 $.
The solid line shows $Q$ before projection at scale $ k=\kappa/\Dstar $, 
multiplied by the factor in \protect{\eq{QGP}}.
\label{figQkappa}}

\end{document}